\documentclass [prl, superscriptaddress, showpacs, twocolumn ]{revtex4}
\usepackage{graphicx}
\usepackage{dcolumn}

\newcommand{\YBa}{YBaCo$_4$O$_7$}
\newcommand{\REBa}{RBaCo$_4$O$_7$}
\newcommand{\Kag}{Kagom\'{e}}
\newcommand{\YCaBa}{Y$_{0.5}$Ca$_{0.5}$BaCo$_4$O$_7$}
\begin{document}

\title{Magnetic correlations in YBaCo$_4$O$_7$ probed by single-crystal neutron scattering}

\author{P. Manuel}
\email{pascal.manuel@stfc.ac.uk}
\affiliation{ISIS facility, STFC Rutherford Appleton
Laboratory, Chilton, Didcot, Oxfordshire, OX11 0QX,
United Kingdom.}
\author{L. C. Chapon}
\affiliation{ISIS facility, STFC Rutherford Appleton
Laboratory, Chilton, Didcot, Oxfordshire, OX11 0QX,
United Kingdom.}
\author{P.G. Radaelli}
\affiliation{ISIS facility, STFC Rutherford Appleton
Laboratory, Chilton, Didcot, Oxfordshire, OX11 0QX,
United Kingdom.}
\author{H. Zheng}
\affiliation{Materials Science Division, Argonne National Laboratory, Argonne, Illinois, 60439, USA}
\author{J.F. Mitchell}
\affiliation{Materials Science Division, Argonne National Laboratory, Argonne, Illinois, 60439, USA}

\date{\today}

\begin{abstract}
We have studied the frustrated system YBaCo$_4$O$_{7.0}$ generally described as an alternating stacking of \Kag\ and triangular layers of magnetic ions on a trigonal lattice, by single crystal neutron diffraction experiments above the N\'{e}el ordering transition. Experimental data reveals pronounced magnetic diffuse scattering, which is successfully modeled by direct Monte-Carlo simulations. Long-range magnetic correlations are found along the \textit{c}-axis, due to the presence of corner-sharing bipyramids, creating quasi one-dimensional order at finite temperature. In contrast, in the \Kag\ layers (\textit{ab}-plane), the spin-spin correlation function -displaying a short-range 120$^{\circ}$ configuration- decays rapidly as typically found in spin-liquids. \YBa\ experimentally realizes a new class of two-dimensional frustrated systems where the strong out-of-plane coupling does not lift the in-plane degeneracy, but instead act as an external "field".
\end{abstract}


\maketitle

\begin{figure}[h!] 
\includegraphics[width=210pt]{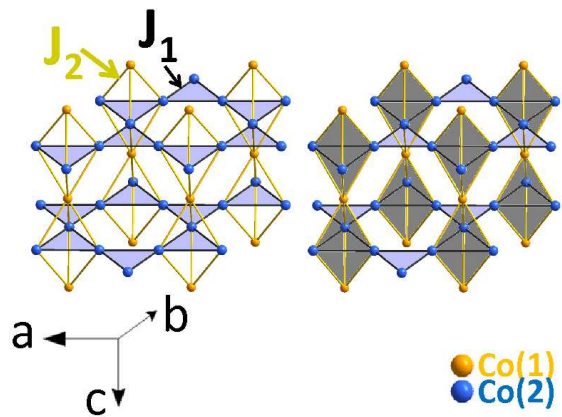} 
\caption{(Color online) Sketch of the magnetic topology in the \YBa\ compound. Orange and blue spheres represent the inequivalent Co sites in, respectively, the Co(2) \Kag\ and Co(1) triangular layers (see text for details). The triangles in the \Kag\ planes are colored in blue for clarity (left panel) and bipyramids are colored in dark grey (right panel). Black and orange lines represent magnetic exchange interactions J$_1$ and J$_2$}
\label{fig:structure}
\end{figure}

\begin{figure}[h!] 
\includegraphics[scale=0.45]{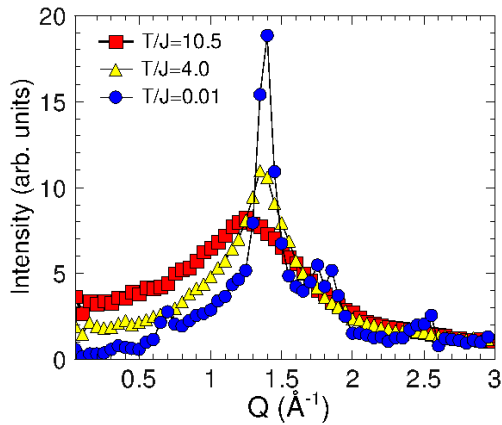} 
\caption{(Color online) Simulated magnetic scattering cross-section for a polycrystalline sample of \YBa\ at various temperatures. The powder-averaged intensities have been calculated by numerical integration of the calculated single-crystal scattering cross sections (see text for details).}
  \label{fig:powder}
\end{figure}

\indent Geometrically frustrated (GF) magnets continue to attract both experimental and theoretical attention due to the fact that exotic magnetic states such as spin-ices and spin-liquids emerge from frustrated magnetic interactions. In such systems, the impossibility of uniquely minimizing all energy terms in the magnetic Hamiltonian, leads to a macroscopically degenerate manifold of ground states, which in the absence of perturbations, prevents the formation of long-range magnetic order. A classic example of GF systems extensively studied theoretically \cite{MoessnerChalker} include vertex-sharing triangular or tetrahedral networks with antiferromagnetic (AFM) interactions, realized experimentally in, for instance, SrCr$_8$Ga$_4$O$_{19}$ \cite{SCGO}.  \\
\indent The recently discovered class of compounds \REBa\ (R=rare-earth, Y) has attracted considerable attention \cite{ISI:000241855500005,schweika:067201} since it is among the few crystalline solids whose structural motif embraces magnetic \Kag\ layers, a classical GF lattice. The transition metal magnetic sublattice of \REBa\ is made of triangular and \Kag\ layers of Co ions (Co(1) and Co(2), respectively) that are alternatively stacked along the \textit{c}-axis of the trigonal crystal structure, as shown in Fig.\ref{fig:structure}. Strong AFM interactions are mediated through Co-O-Co superexchange pathways between first neighbor Co ions, tetrahedrally coordinated by oxygens, creating an exchange topology based on triangular motifs. Magnetic susceptibility measurements on all analogs of the \REBa\ family are characteristic of strongly frustrated systems with a building up of strong correlations much above the ordering temperature T$_N$ as shown by the deviation of the susceptiblity from linearity below the Weiss temperature. In the specific case of R=Y, oxygen stoichiometric samples eventually order magnetically at low temperature (T$_N$=110K) due to structural distortion at T$_S$ $>$ T$_N$ that lifts the three-fold degeneracy and $\theta_{CW}$=-508K\cite{ISI:000242409000015}. YBaCo$_4$O$_{7.0}$ is thus a moderately GF magnet (f=$\theta_{CW}$/T$_N$ $\sim$ 5). Interestingly, oxygen-rich samples retain three-fold symmetry and a disordered ground state down to the lowest temperature\cite{ISI:000236501300023}.\\ 
\indent Although several neutron scattering studies have been recently conducted, the exact nature of the magnetic ground state in these systems remains unclear. Soda et al.\cite{ISI:000237969600046} have reported two magnetic transitions in single crystal of \YBa\, both with short-range correlations, suggesting that the magnetic behavior of the triangular and \Kag\ layers are essentially decoupled. On the other hand, our powder diffraction experiments \cite{ISI:000242409000015} revealed that long-range magnetic order appears at a single transition (T$_N$=110K) for oxygen-stoichiometric sample of \YBa. The magnetic structure derived from Rietveld refinement, shows that both magnetic Co sites bear a significant ordered moment, with a complex non-colinear magnetic configuration, contradicting the picture of decoupled layers.\\
\indent  Using single crystal neutron diffraction combined with Monte-Carlo (MC) simulations of the magnetic diffuse scattering collected at $\Theta_{CW}>$T=130K$>T_N$, we demonstrate that \YBa\ represents a new GF system, characterized by the coexistence of long-range quasi-one dimensional (1D) correlations and strong degeneracy in the two remaining directions. Our analysis reveals that the \emph{primitive} magnetic clusters are corner-sharing triangular and bipyramidal units, and minimization of the intra-cluster exchange energy dictates configurations with zero total moment in each cluster. The form of the spin Hamiltonian uniquely promotes strong 1D spin-spin correlations between apical Co(1) sites within chains running along the c-axis. This contrasts with the fastly decaying correlations along any direction in the \textit{ab}-plane, characteristic of spin-liquid with 120$^{\circ}$ correlations. Our model provides an alternative explanation to that of Schweika et al. \cite{schweika:067201} who describe the magnetic scattering from polycrystalline \YCaBa\ as arising from staggered chirality and for which the Co(1) site is non-magnetic.\\
\indent A single crystal of YBaCo$_4$O$_{7.0}$ was grown by a floating zone technique in an optical image furnace. A densified rod of the nominal composition was melted in a 20\%\ O$_2$/Ar atmosphere, and the crystal grown at 1 mm/hr. Because of the high affinity for YBaCo$_{4}$O$_{7}$ to pick up oxygen on cooling \cite{ISI:000236501300023, ISI:000250971200035, ISI:000257281900036}, an after-heater was used to keep the growing crystal above 900$^{\circ}$C. After growth was complete, the atmosphere was changed to 99.995 \%\ Ar for 24 hours and the crystal cooled down to room temperature over six hours. Oxygen content was verified by thermogravimetric analysis of a crushed sample of the crystal. Diffuse magnetic neutron scattering maps were recorded on the PRISMA time-of-flight spectrometer at the ISIS facility, Rutherford Appleton Laboratory (UK). 
A $\sim$ 200 mg piece of the single crystal of cylindrical shape (5 mm diameter, 2 mm height) was mounted on an aluminium pin with the [001] axis and [100] aligned vertically, in turn. For each axial orientation, scattering in the orthogonal reciprocal-space planes were recorded by rotating the crystal around the vertical axis. Data have been corrected for absorption and normalized to the incoherent scattering of a vanadium standard. The disordered magnetic arrangements have been derived from direct Monte-Carlo (MC) simulations \cite{diffuseprogram} using a standard Metropolis algorithm \cite{metropolis} using Heisenberg spins and isotropic exchange interactions. Simulations have been performed using a supercell made of 12x12x12 conventional hexagonal unit-cells and cyclic exchange boundary conditions, giving a total of 13824 spins. Magnetic neutron scattering intensities have been obtained by calculating the magnetic structure factor of the supercell \footnote{The intensity for a given scattering vector \textbf{Q} is I(\textbf{Q})=$\Vert\widehat{\mathbf{Q}}\times\mathbf{F_M(Q)}\times\widehat{\mathbf{Q}}\Vert^2$ where $\widehat{\mathbf{Q}}$ is a unit vector in the direction of \textbf{Q} and the unit-cell magnetic structure-factor $\mathbf{F_M(Q)}=\frac{\gamma r_0}{2N_xN_yN_z}\sum_i f_i(\mathbf{Q})\mathbf{M_i} e^{i\mathbf{Q.R_i}}$ where $\gamma$ is the neutron gyromagnetic ratio, $r_0$ the free electron radius, $N_x$, $N_y$, $N_z$ are the supercell dimensions. The sum runs other all magnetic sites in the supercell, of position $R_i$ and moment $M_i$ and f$_i$(\textbf{Q}) is the magnetic form factor for Co$^{2+}$}. Powder-average was calculated by integrating the single-crystal scattering cross-sections in a hemisphere choosing discrete Q points that were homogeneously distributed.\\
\indent Neutron diffraction data collected at 330K are consistent with the trigonal space group $P31c$ already proposed from measurements on powder samples \cite{ISI:000242409000015}. On cooling below T$_S$=300K, superlattice reflections corresponding to the orthorhombic $Pbn2_1$ symmetry \cite{ISI:000242409000015,ISI:000236501300023}, not shown, are observed for the three 120$^{\circ}$ domains.  However for the sake of simplicity, we assume that the system can be modelled by a minimal set of exchange interactions on an ideal trigonal structure. There are four inequivalent Co-O-Co superexchange integrals to consider in this geometry, two in plane, between nearest neighbor Co(2) (non-capped triangle and capped triangle of the \Kag\ planes respectively), and two out of plane, between Co(2) and Co(1) (triangular layer above the \Kag\ layer or below). A spin-dimer analysis using the Extended-Huckel Tight-Binding method \cite{khalyavin} shows that the two in-plane and two out-of-plane interactions do not differ by more than 5\%. We have therefore assumed only two inequivalent interactions J$_1$ and J$_2$, as shown in Fig.\ref{fig:structure}. 
\begin{figure}[h!] 
\includegraphics[width=220pt]{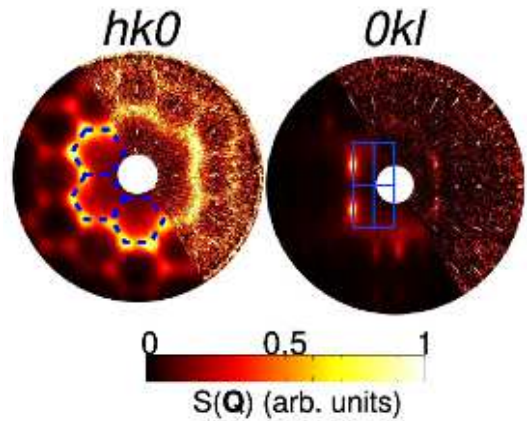} 
\caption{(Color online) Magnetic diffuse scattering for \YBa\ in the hk0 (left panel) and 0kl (right panel) reciprocal planes. Each map shows the scattering cross sections obtained from Monte-Carlo simulations (left hand side) and the experimental magnetic scattering (right hand side) derived by subtracting the nuclear scattering collected at 330K from the total scattering at 130K. Dashed blue lines indicate the Brillouin zones (left) while solid lines (right) show the reciprocal lattice.}
\label{fig:diffusemap}
\end{figure}
Fig.\ref{fig:diffusemap} shows the magnetic scattering in the hk0 and 0kl planes, obtained by subtracting the nuclear contribution (330K data) from the data collected at 130K, just above T$_N$. Although this procedure only partially removes the nuclear scattering, as evidenced by the presence of residual sharp scattering at Bragg positions, it clearly highlights the magnetic signal. The latter is not resolution limited and rapidly decays with the norm of the scattering vector ($\vert \mathbf{Q} \vert$), as expected from the magnetic form factor dependence. The in-plane magnetic scattering is relatively diffuse and concentrated along certain edges of the hexagonal Brillouin Zones (BZ). Remarkably, there is very little contribution in the first BZ and most of the intensity is found at the external perimeter of the second BZs. Reciprocal space cuts along directions perpendicular to the diffuse features leads to a full width at half maximum (FWHM) of 0.1$\AA^{-1}$, i.e. correlation lengths of about 30$\AA$. The nature of the diffuse scattering in the 0kl plane is characterized by isolated diffuse spots at reciprocal space positions 0,$\pm\frac{1}{2}$,2, the most intense features, and a spot a 0,$\frac{3}{2}$,0. The first set of peaks display anisotropic broadening with a FWHM along b$^*$ of 0.26$\AA^{-1}$ about 2 times larger than the FWHM along c$^*$ of 0.12$\AA^{-1}$, indicative of much longer correlations along the \textit{c}-axis.\\
\indent MC simulations have been performed for a wide range of J$_2$/J$_1$ ratios \cite{khalyavin}. The resulting phase diagram will be the object of a separate publication. It is found that the experimental magnetic scattering maps presented here are best reproduced by a value of J$_2$/J$_1$ close to one. The corresponding calculated magnetic scattering is plotted alongside the experimental data on Fig. \ref{fig:diffusemap}. These results were obtained by simulated annealing from a temperature T/J=100 down to a chosen final T/J. The maps in Fig.\ref{fig:diffusemap}, obtained by averaging the results of ten MC runs, are shown for T/J=4.0 which provides the best agreement with the experiment. This value is close to the calculated T/J=3.75 at T=130K if one take into account the mean-field AFM exchange interaction ($\langle J \rangle\sim$ 35K) extracted from the paramagnetic susceptibility \cite{ISI:000242409000015}. The agreement with the data is excellent, reproducing all main features, i.e. positions and relative intensities, in both scattering planes establishing that the essential physics is captured by our simple model. 
\begin{figure}[h!]
\includegraphics[width=240pt]{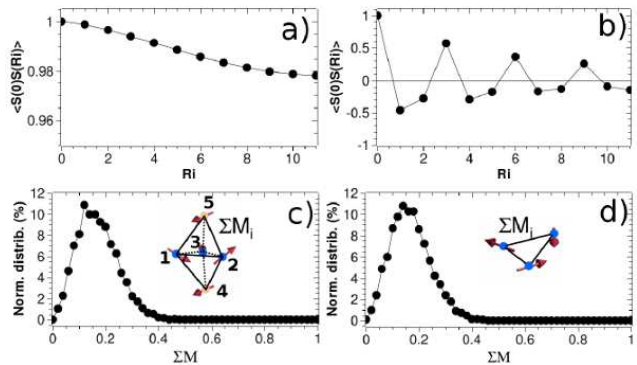} 
\caption{(Color online) Static spin-spin correlation functions for the Co(1) site along the z-axis (panel a) and in-plane ([100],[010], or [1-10]-directions, panel b). Distributions of the magnetization within triangular units of the \Kag\ lattice (panel c) and within bipyramidal units (panel d). Sketches of the triangular and bipyramidal units together with typical spin configurations are depicted in the insets.}
\label{fig:correlations}
\end{figure}
\indent Crucial insights into the magnetic behavior of \YBa\ are unveiled by parameters extracted from the zero-temperature ground state of the MC simulations. For example, analysis of the spin configurations, reveals that the magnetic lattice is made of triangular and bipyramid units, both with non-colinear arrangements and satisfying the $\mathbf{\sigma}$ =$\sum_i \mathbf{S_i}$=0 rule. The latter point is illustrated in Fig.\ref{fig:correlations} where the distribution of the total moment $\sigma$  taken over the entire lattice, shown for triangular and bipyramidal units, is centered around very small values. The small deviations from zero are related to inevitable residual fluctuations in the MC algorithm, frozen at low temperature. We also found that the magnetic moments on the two apical positions of any bipyramid are systematically aligned parallel, a point developed in the next section.\\
\indent The $\mathbf{\sigma}$=0 rule must be satisfied in any cluster of mutually connected spins \cite{frustrated}, such as for the triangular units imposing a angle of 120$^{\circ}$ between spins. However, spins in the trigonal bipyramid are not all mutually connected, since no nearest-neighbor exchange paths mediate interactions between the apical sites labeled 4 and 5 in the inset of Fig.\ref{fig:correlations}. However, for a unique exchange integral J=J$_1$=J$_2$, one can rewrite the Hamiltonian of the bipyramid up to a constant in a simple way:
\begin{equation}
H=\frac{J}{2} \left[ \left( \sum_{i=1}^{5} \mathbf{S_i} \right)^2 - 2 \mathbf{S_4} \cdot \mathbf{S_5} \right]
\label{eq:bipyramidH}
\end{equation}  
This form also imposes $\sigma$=0 to minimize the energy. Additionally, Eq.\ref{eq:bipyramidH} indicates that only parallel alignment of S$_4$ and S$_5$ are possible ground states, in agreement with the configurations found in the MC simulations. This point is reinforced by examination of the static spin-spin correlation (SSC) function along the \textit{c}-axis. The SSC for a given magnetic site S$_i$ in a direction \textbf{R} is defined as:
\begin{equation}
\langle S(0) S(R) \rangle = \sum_i \mathbf{S_i} \cdot \mathbf{S_{i+R}}
\end{equation} 
where the sum runs over all spins i in the lattice. In order to access spatial correlations at long distances in all directions, a MC simulation with a 24x24x12 supercell was performed. The SSC functions have been calculated for T/J=0.01 and are presented in Fig.\ref{fig:correlations}. The SSC functions along the \textit{c}-axis decay extremely slowly. This is illustrated for the apical Co(1) site, with 98\%\ correlations at the tenth neighbor. This is consistent with Eq. 1 since the bipyramids are corner-sharing along the \textit{c}-axis and explains why the diffuse spots in the 0kl map are sharper along c$^{*}$. In contrast, correlations between Co(1) in-plane are lost much faster(25\%\ at the ninth neighbor). They are also reminiscent of a 120$^{\circ}$-configuration, characterized by the oscillating nature of the SSC with a periodicity R=3 and a value of -0.5 for R=1. Again, this behavior agrees  with the topology of the lattice since first neighbor bipyramids in the \textit{ab}-plane are connected through a triangular unit that imposes such 120$^{\circ}$ configuration.\\
\indent Further insight into the nature of the spin texture in \YBa\, is provided by the correlations of both the vector  ($\mathbf{K_v}$) and scalar ($\mathbf{K_s}$) chiralities for triangular units, as defined in \cite{chirality}. Indeed, in contrast to perfect \Kag\ lattices with staggered (q=$\surd$3$\times\surd$3 ) or uniform (q=0) $\mathbf{K_v}$, or to the non-zero $\mathbf{K_s}$ in the field-induced spin-canted structure seen in \Kag\ iron jarosite \cite{fejarosite}, our model doesn't exhibit any sign of such correlations.\\
\indent Finally, one can derive the powder-averaged magnetic cross section (Fig.\ref{fig:powder}) for several values of T/J including that which most closely matches our single crystal data. 
While the extracted intensity agrees with our neutron powder data for \YBa\ \cite{ISI:000242409000015}, with the presence of a main peak at Q=1.35$\AA^{-1}$, it also closely matches that observed in \YCaBa\  \cite{schweika:067201}, where the apical Co ions are reported to be non-magnetic. In that work, Schweika et al described the correlations in the latter system as arising from a q=$\surd$3$\times\surd$3 structure, in which local zero-energy thermal weathervane defects, on the scale of one hexagon of the \Kag\ motif, results in an intense peak at 2q$\surd_3$=1.35$\AA^{-1}$ but breaks longer range correlations, hence the weak signal they observed at q$\surd_3$=0.78$\AA^{-1}$. In our model, the peak at 0.78 $\AA^{-1}$ is naturally very weak near zero-temperature and is non existent at slightly higher ratios of T/J due to the weak scattering within the first BZ.\\ 
\indent In summary, our magnetic neutron diffraction study in the spin-liquid regime of the frustrated \YBa\ system, reveals that the magnetic lattice is best described as a 3D network of corner-sharing triangles and trigonal bipyramidal clusters, as illustrated on the right panel of Fig.\ref{fig:structure}. Magnetic correlations, long-range between apical Co ions along the \textit{c}-axis and short-range in the \textit{ab}-plane, preserving 120$^{\circ}$ configuration, bear the signature of a new type of magnetic lattice. Despite the strong out-of-plane coupling between the \Kag\ layers in \YBa\ acting as a field and constraining the spin configurations for half of the triangular plaquettes, the system retains a strong in-plane degeneracy. This unique feature should be relevant to explain physical properties for other analogs of the series and will no doubt spur experimental work in the samples where the three-fold symmetry remains at low temperatures.\\
\indent The authors would like to thank John Chalker for stimulating discussions.

\end{document}